\documentclass{aastex}
\usepackage{emulateapj5}
\shorttitle{A probe of solar interior magnetic fields}
\newcommand{\p}[2]{\frac{\partial #1}{\partial #2}}

\newcommand{\od}[2]{\frac{d #1}{d #2}}

\begin{document}

\title{Measurements of solar irradiance and effective temperature as a probe of solar interior magnetic fields}
 
\author{L. H. Li\altaffilmark{1,2} and S. Sofia}
\affil{Department of Astronomy, Yale University, P.O. Box 208101, New Haven, CT 06520-8101}
\altaffiltext{1}{also at Purple Mountain Observatory,  Chinese Academy of Sciences}
\altaffiltext{2}{also at National Astronomical Observatories, Chinese Academy of Sciences}
\email{li@astro.yale.edu}
\email{sofia@astro.yale.edu}

\begin{abstract}
We argue that a variety of solar data suggest that the activity-cycle
timescale variability of the total irradiance, is produced by structural
adjustments of the solar interior. Assuming these adjustments
are induced by variations of internal magnetic fields, we use measurements
of the total irradiance and effective temperature over the period from 1978
to 1992, to infer the magnitude and location of the magnetic field. Using
an updated stellar evolution model, which includes magnetic fields,
we find that the observations can be explained by fields whose peak values
range from 120k to 2.3k gauss, located in the convection zone between
$0.959R_{\sun}$ and $0.997R_{\sun}$, respectively. The corresponding maximal
radius changes, are 17 km when the magnetic field is located at $0.959R_{\sun}$
and 3 km when it is located at $0.997R_{\sun}$. At these depths, the $W$ parameter
(defined by $\Delta \ln R / \Delta \ln L$, where $R$ and $L$ are the radius and 
luminosity) ranges from $0.02$ to $0.006$. All these predictions are consistent with
helioseismology and recent measurements carried out by the MDI experiment on
SOHO.
\end{abstract}

\section{Introduction}\label{s1}

Direct satellite measurements of total solar irradiance \cite{WH91, FL98},
show that it varies almost in phase with the solar activity cycle, and that
its relative variation in amplitude is about 0.1\%. Although the most
common explanation of this change, involves the effect of magnetic
regions and network (superposed to an otherwise constant solar photosphere),
an alternative explanation is that the cyclic variation is primarily due to
structural changes of the solar interior. Two arguments support this second
possibility. Firstly, direct measurements of the spectrum of p-mode
oscillations, show variations \cite{LW90} that can be best explained in terms
of structural changes  (i.e. changes in pressure, density, radius, etc)
\cite{LGS96,ACT00,BGM96}. Secondly, measurements of the ``effective
temperature'' by Gray and Livingston (1997b), also vary  nearly in
phase with the solar cycle. This variation is obtained by monitoring the
neutral carbon $\lambda5380$ line in the solar flux spectrum. They present
several arguments supporting their contention that what they measure is the
gas temperature in the deep photosphere, free from the influence of
sunspots, faculae and small scale magnetic flux tubes. Their main argument
rests on their simultaneous monitoring of C I $\lambda5380$, Fe I
$\lambda5379$ and Ti II $\lambda5381$. They use the ratios of spectral line
depths, C I $\lambda5380$ to Fe I $\lambda5379$ and to Ti II $\lambda5381$,
as temperature indicators. Since these three lines have different excitation
potentials ($1.57$ eV for Ti II $\lambda5381$, $3.69$ eV for Fe I
$\lambda5379$, and $7.68$ eV for C I $\lambda5380$), the consistency of the
results for the two line depth ratios, shows that the temperature
indicators are free from the influence of variations of faculae and small scale
magnetic flux tubes, which would be expected to affect each line in a
different way. See Gray \&\ Livinston (1997a,b) for the entire discussion.
In order for the gas temperature to change, the energy flow from the
interior must also change. Structural adjustment will be able to accomplish
this. Since the relative radius variation $\Delta\ln R$ is very 
small \cite{EKBS00},
\begin{equation}
  \Delta\ln L \approx 4\Delta\ln T,
\end{equation}
where $L$ and $T$ are luminosity and temperature, respectively.
The temperature change measured by Gray and Livingston (1997) is $1.5\pm
0.2$ K in one cycle, which corresponds to $\Delta\ln T\sim 0.025\%$ and 
$\Delta\ln L\sim 0.1\%$. This is approximately the entire cyclic variation
of the total irradiance.

\vspace{3mm}
\centerline{\epsfysize=5.5cm \epsfbox{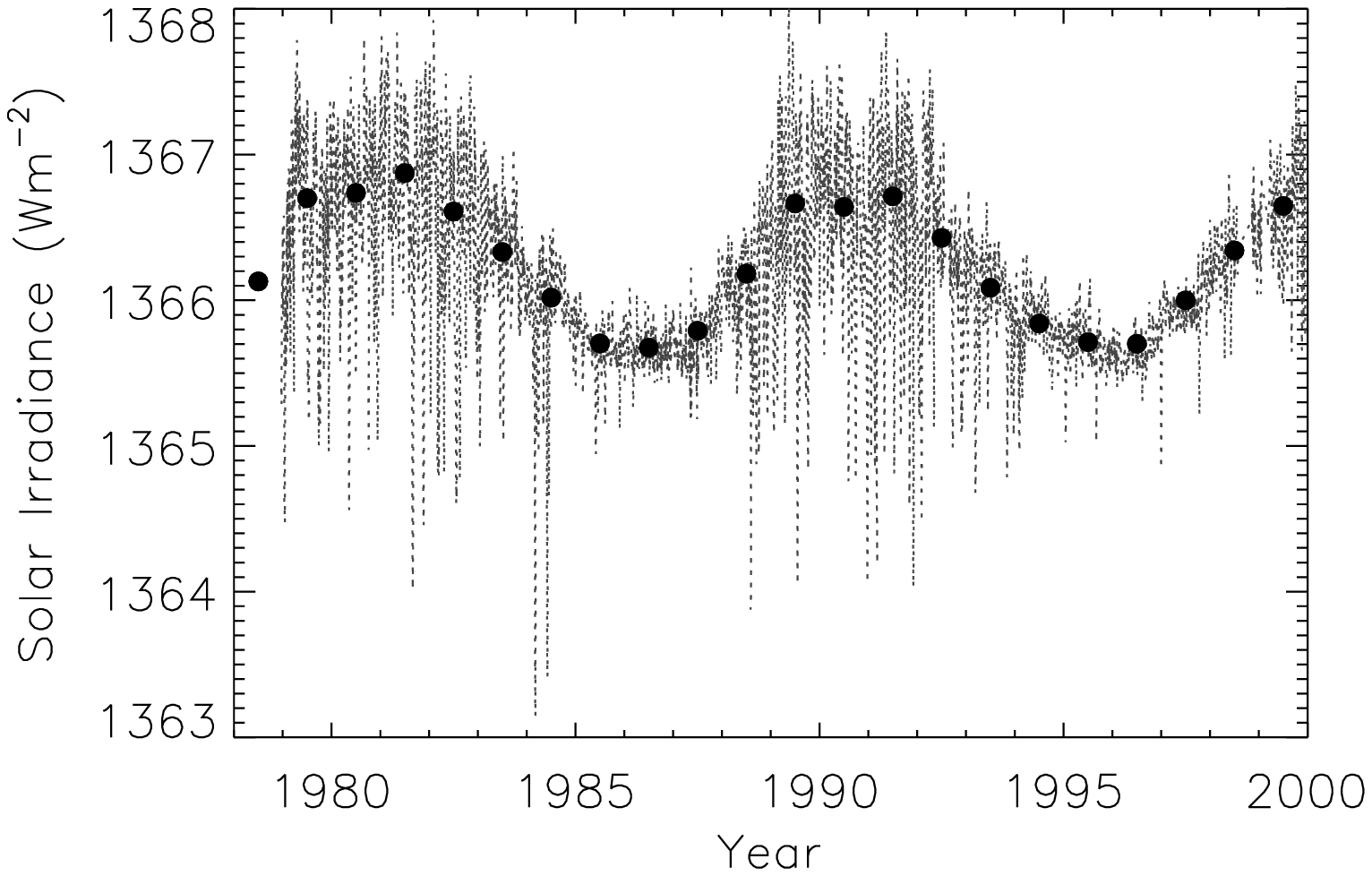}}
\figcaption[lifig1.eps]{
A composite solar irradiance record from the end of 1978 to the present
\protect\cite{FL98} and the yearly mean of solar irradiance.
\label{lifig1}
}
\vspace{3mm}

The most obvious way the solar structure may change, is in response to
variations of the internal magnetic field during the dynamo build-up and
decay. In order to compute this effect, we introduce magnetic variables as
new stellar structure variables \cite{LS95}, in addition to the conventional
ones \cite{GDKP92}. Since magnetic field is not a scalar, we have to use at
least two variables to describe it: the magnetic energy density $\chi$
\cite{LS95}, and the ratio of the magnetic pressure to the magnetic energy,
$\gamma-1$ \cite{EST85}. This method was developed by Lydon and Sofia in
1995, in which $\gamma$ was treated as a parameter, and it was subsequently
applied by Lydon, Guenther and Sofia (1996), to explain the observed
variation of solar p-mode oscillations with the solar cycle \cite{LW90}.
Here we want to update this method by: (i) treating both $\chi$ and $\gamma$
as new structure variables, as done for $\chi$ in Lydon and Sofia (1995),
(ii) taking into account the influence of magnetic fields on radiative
opacities, (iii) taking into account all time-dependent contributions to the
equations of stellar structure (since we want to treat short time scales),
(iv) modifying the radiative loss assumption of a convective element (as
discussed in section 3), (v) updating the code by using the new version of
Yale Stellar Evolution Code (YREC7) and (vi) removing the
perfect gas approximation, which was assumed when calculating some first- and
second-order derivatives associated with magnetic fields in Lydon and Sofia 
(1995). Because luminosity, radius and temperature variations are sensitive 
to the location, intensity, orientation, and distribution of the perturbation
magnetic field $B=(8\pi \chi\rho)^{1/2}$ ($\rho$ is the gas density), we can 
use the measured yearly-averaged irradiance, temperature and radius variations,
to determine the solar interior magnetic field as a function of mass and time.

\vspace{3mm}
\centerline{\epsfysize=5.5cm \epsfbox{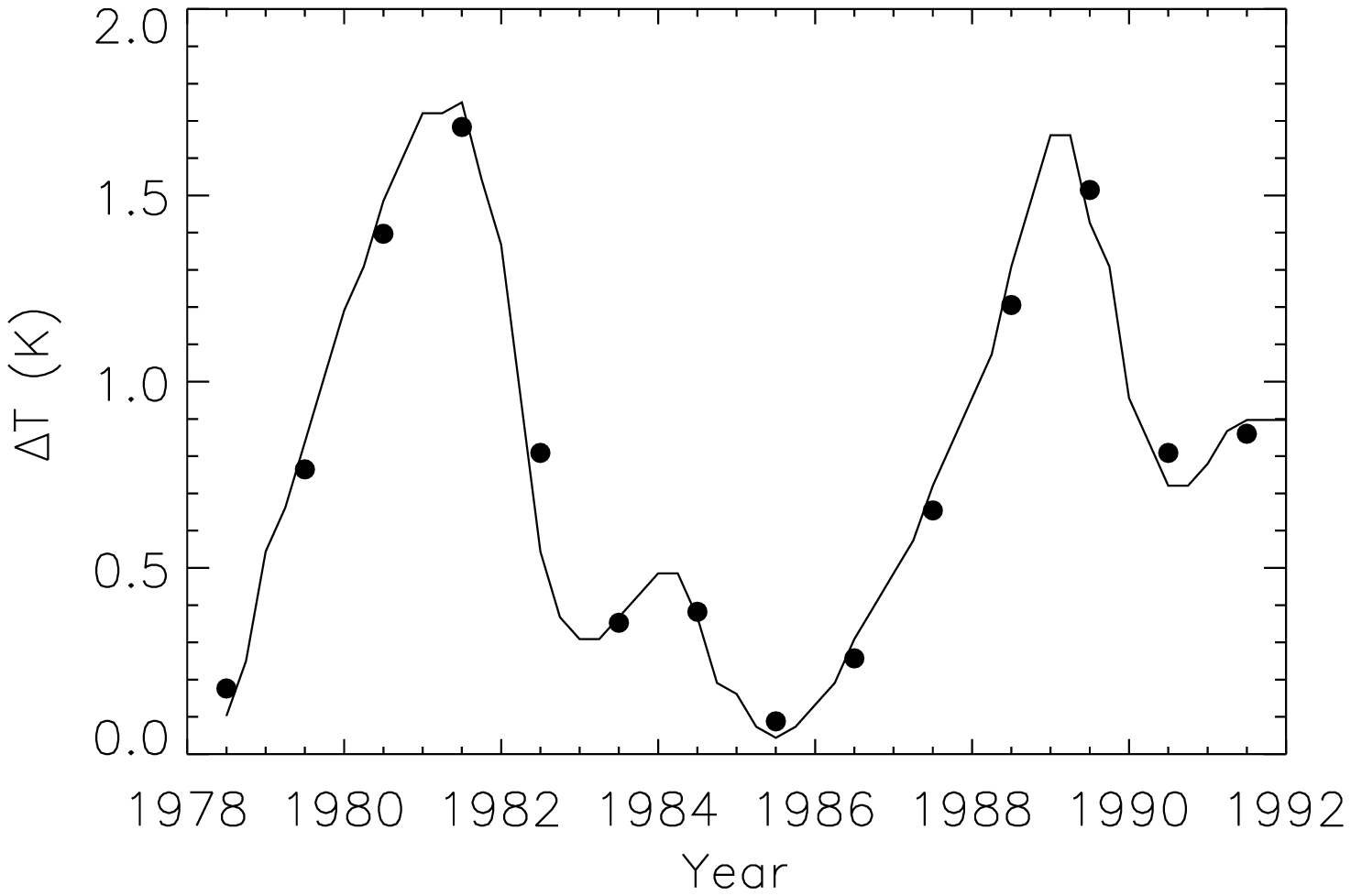}}
\figcaption[lifig2.eps]{
The measured solar photospheric temperature variations from 1978 to 1992
\protect\cite{GL97b} and the yearly mean.
\label{lifig2}
}
\vspace{3mm}

\section{Solar irradiance and effective temperature records}

Space-based solar irradiance measurements have been reviewed by Fr\"ohlich
and Lean (1998), and a composite solar irradiance record for the period from
1978 to the present can be downloaded from the website
http://www.ngdc.noaa.gov/. In our magnetic perturbation model calculations,
we use one year as the timestep, since variations on the shorter time scales
may be attributed to the effect of magnetic regions and network. We
compare the results with the yearly means of total solar irradiance, as
shown in Figure~\ref{lifig1}. The standard variances range from 0.1
Wm$^{-2}$ to 0.5 Wm$^{-2}$.

The spectroscopic temperature variations of the sun, measured by Gray and
Livingston (1997) over the period from 1978 to 1992, are shown in
Figure~\ref{lifig2} (reproduced from their Figure 9). The points are
the yearly means, the zero point being chosen arbitrarily.

\section{Method}

\subsection{Definition of variables}

Standard solar models (SSM) use pressure, temperature, radius and luminosity
as the structure variables, and mass $M_r$ interior to a radius $r$, as
the independent variable. In order to self-consistently take into account
magnetic effects, Lydon and Sofia (1995) introduced two new magnetic
structure variables to model the sun: the magnetic energy per unit mass
$\chi$, and the effective ratio of specific heats for the magnetic
perturbation, $\gamma$. The former describes the magnetic perturbation
strength, and the latter describes the tensor feature of the magnetic
pressure. In general, the determination of $\chi$ and $\gamma$ requires a
comprehensive understanding of turbulent dynamics in the solar convective
zone \cite{BCT95}, an undertaking that is impractical at present. Therefore
we take $\chi$ and $\gamma$ as free parameters (functions) that need to be
determined at this stage, instead of using dynamo equations to determine
them.

Our aim is to use the measured solar irradiance and effective temperature
variations given in Figures~\ref{lifig1} and \ref{lifig2} to determine
$\chi(M_D,t)$ and $\gamma(M_D,t)$, where $M_D$ and
$t$ are mass depth and time, respectively.
Using  $\chi(M_D,t)$, we can obtain the  solar interior magnetic field
$B(M_D,t)=(8\pi \chi \rho)^{1/2}$. Since these observations are not enough
to determine the two new variables uniquely, we assume they have a gaussian-like
shape, and also utilize helioseismic (\cite{ACT00}) and radius
(\cite{EKBS00}) observations, as additional constraints. The mass depth $M_D$
is defined as
\begin{equation}
  M_D = \log_{10}(1-M_r/M_{\sun}) \label{md}.
\end{equation}
This replaces the mass variable $M_r$ when we want to describe the location,
orientation and form of all magnetic perturbations.

For our purpose we can use the toroidal ($B_t$) and poloidal ($B_p$)
components to express a magnetic field vector
$\vec{B} = (B_t,B_p)$. The magnetic energy density
variable $\chi$ can be written as,
\begin{equation}
\chi =  (B^2/8\pi)/\rho,
\end{equation}
where $B=(B_t^2+B_p^2)^{1/2}$ is the magnitude of the magnetic field vector.
The magnetic field direction variable $\gamma$ can be written as,
\begin{equation}
  \gamma = (2B_t^2+B_p^2)/B^2.
\end{equation}
The magnetic pressure $P_{\chi}$ can be defined in terms of the new magnetic
variables as
\begin{equation}
  P_\chi = (\gamma-1)\chi \rho.
\end{equation}
The value of $\gamma$ depends on the field geometry. Parallel to the field
lines, $P_\chi$ is zero, so $\gamma=1$, whereas perpendicular to the field 
lines, $P_\chi$ is a maximum, so $\gamma=2$.

\subsection{Consequences on the equations of stellar structure}

The continuity equation remains the same. The hydrostatic equation is replaced
by the equation of motion, which including the Lorentz force, is
\[
  \rho \frac{\partial^2\vec{r}}{\partial t^2} = -\nabla P -
\frac{GM_r}{r^2}\rho \hat{r} + \frac{1}{4\pi}
    (\nabla\times\vec{B})\times\vec{B},
\]
where $G$ is the gravitational constant. The radial component is
\[
   \rho \frac{\partial^2r}{\partial t^2} = - \frac{\partial P}{\partial r} -
\frac{GM_r}{r^2}\rho - \p{P_{\chi}}{r}.
\]
The magnetic field has both a magnetic tension and a magnetic pressure.
This can be seen by writing the Lorentz force
$(1/4\pi)(\nabla\times\vec{B})\times\vec{B}$, as the sum of the magnetic
tension $(\vec{B}\cdot\nabla)\vec{B}/4\pi$ and the magnetic pressure
$-\nabla(B^2/8\pi)$. The former is anisotropic, and the latter is isotropic.
Both contribute to the last term of the radial motion equation, and are 
incorporated into $P_{\chi}$. For example, for a purely radial magnetic field,
the magnetic tension force, $(1/4\pi)B_r\partial B_r/\partial r=(1/8\pi)\partial
B_r^2/\partial r=(1/8\pi)\partial B^2/\partial r$, which cancels the
magnetic pressure force. As a result, $P_{\chi}=0$ in this case.

Defining
\[
   P_T=P+P_{\chi}
\]
as the total pressure, the equation of radial motion becomes
\begin{equation}
  \p{P_T}{M_r} = - \frac{GM_r}{4\pi r^4} - \frac{1}{4\pi
r^2}\frac{\partial^2{r}}{\partial{t^2}}, \label{eq:motion}
\end{equation}
where the last term represents the inertial force. Since
\[
  \rho = \rho(P_T, T, \chi, \gamma),
\]
the equation of state becomes
\[
  \frac{d\rho}{\rho} = \alpha \frac{d P_T}{P_T} - \delta \frac{d T}{T} -\nu
\frac{d\chi}{\chi} - \nu'\frac{d\gamma}{\gamma},
\]
where
\begin{eqnarray}
\alpha &=& \p{\ln\rho}{\ln P_T} \mbox{ at constant T, $\chi$, $\gamma$}
\nonumber \\
\delta &=& -\p{\ln\rho}{\ln T} \mbox{ at constant $P_T$, $\chi$, $\gamma$}
\nonumber \\
\nu &=& -\p{\ln\rho}{\ln \chi} \mbox{ at constant $P_T$, T, $\gamma$}
\nonumber \\
\nu' &=& -\p{\ln\rho}{\ln \gamma} \mbox{ at constant $P_T$, T, $\chi$}
\nonumber
\end{eqnarray}
Since we use the one-dimensional stellar evolution model, we
can not model the transverse component of the equation of motion. In
order to handle this transverse component, we need at least a two-dimensional
stellar evolution model. Since such a model is not available yet, we only 
consider the radial motion equation in this paper.

The energy conservation equation becomes
\begin{equation}
  \p{L}{M_r} = \epsilon - T\od{S_T}{t} - \frac{1}{\rho}\p{u}{t}.
\label{eq:energy}
\end{equation}
Here
\begin{eqnarray}
  TdS_T &=& dQ_T = dU + PdV + d\chi \nonumber \\
             &=& dU_T + P_TdV - (\gamma -1)(\chi/V) dV \nonumber \\
             &=& c_{\rm p}dT - \frac{\delta}{\rho} dP_T + \left(1+\frac{P_T\delta
\nu}{\rho\alpha\chi}\right)d\chi +
                       \frac{P_T\delta\nu'}{\rho\alpha\gamma}d\gamma
\label{firstlaw}
\end{eqnarray}
is the first law of thermodynamics including magnetic fields.
The total internal energy $U_T = U + \chi$
and the total entropy $S_T=S+\chi/T$. $c_{\rm p}$
is the specific heat at constant pressure. If
$\gamma$ is constant, then $d \gamma = 0$, and equation~(\ref{firstlaw})
reduces to equation~(75) in Lydon \&\ Sofia (1995). The symbol $u=aT^4$,
is the radiation energy density, where $a$ is the
radiation constant.  $\partial u/\partial t$ appears in the full energy 
conservation equation,
\[
    \p{u}{t} + \nabla\cdot \vec{F} = \rho (\epsilon - T\od{S_T}{t}).
\]
$\vec{F} = u\vec{v}$ is the radiation energy flux vector, and $\vec{v}$
is the photon diffusion velocity.

The equation of transport of energy by radiation,
\begin{equation}
  \p{T}{M_r} = - \frac{3}{64 \pi^2 ac}\frac{\kappa l}{r^4T^3},
\label{eq:radiation}
\end{equation}
does not change in form, but the magnetic field affects the opacity
coefficient $\kappa$. Here $c$ is the speed of light. See appendix
\ref{appa} for an estimate of this effect.

The equation of energy transport by convection,
\begin{equation}
  \p{T}{M_r} = - \frac{T}{P}\frac{GM_r}{4\pi r^4} \nabla,
\label{eq:convection}
\end{equation}
does not change form either, but the convection temperature gradient, $\nabla$, 
with a magnetic field, is different from that without it. Lydon and
Sofia (1995) have demonstrated how to account for magnetic effects
in the mixing length theory, and  hence obtain $\nabla$, provided $\gamma$ is
constant.
Their method can also be used to derive $\nabla$ when
$\gamma$ is variable, as we are about to do in this
paper. See appendix~\ref{appb} for the concrete expression for $\nabla$ in
our case.

\subsection{Numerical implementation}

The stellar evolution code solves the stellar structure equations (such
as Eqs.~[\ref{eq:motion}], [\ref{eq:energy}] and  [\ref{eq:radiation}] or
[\ref{eq:convection}], and the continuity equation, $\partial r/\partial M_r=
1/4\pi r^2\rho$), with suitable initial and boundary conditions.
In the model, $M_r$ and $t$ are the independent variables, and
the conventional structure variables are $P_T$, $T$, $r$ and $L$, while the
magnetic variables are $\chi$ and $\gamma$.

To follow the yearly variations of solar global parameters with
a stellar evolution code, requires a timestep no larger than one year.
Such a small timestep requires a very precise code.
For example, because $L_{\sun}$ has increased by about $30\%$
during the lifetime of the sun, the relative mean rate of change is about
$10^{-10}$ per year. To achieve such a high precision is a challenge.
Fortunately, Yale stellar evolution code (YREC) meets this need, as shown
before, \cite{LS95}. We follow Lydon and Sofia (1995) by modifying YREC7, a
new version released in May 1999, in order to accommodate the magnetic
effects described above. The reason why YREC permits a small timestep and
achieves such a high numerical precision, is because it uses
analytical formulae (Prather 1976), rather than numerical methods, to
solve the linearized stellar structure equations.

In order to accommodate the various magnetic effects described above, we
must first write a routine to specify $\chi$ and  $\gamma$. We assume
$\chi=\chi_m(t)F(M_D)$. The maximum magnetic energy density $\chi_m(t)$, 
is to be determined from solar activity indices. The yearly-averaged 
sunspot number, $R_Z$, is the most widely used solar activity index. 
From numerical experiments, we find that the results are sensitive to 
the function form of $B_m$ on $R_z$. If the maximum magnetic field 
in the solar interior, $B_m$, is related to $R_Z$ via
\begin{equation}
  B_m = B_0 \{190 + [1 + \log_{10}(1+R_Z)]^5\}, \label{bm}
\end{equation}
then by adjusting $B_0$, we can nearly match the
measured cyclic variations of irradiance and
effective temperature. The reason why such a functional form of $B_m$ is
chosen is that $B\sim 20$ kG, at a depth of $M_{D}=-4.25$
($r=0.96 R_{\sun}$) in 1996. This result is inferred from helioseismology
\cite{ACT00} when $R_Z$ was at a minimum. Using this prescription for $B_m$,
the value of $B_m$ is about twice as large at the maximum of solar cycle, as
it is at the minimum. 

$F(M_D)$ specifies the distribution of $\chi$, and is to be determined by
fitting the measured irradiance and effective temperature variations.
$F(M_D)$ has infinite degrees of freedom and thus cannot be determined
uniquely by observational results,  which have finite degrees
of freedom. However, we can remove this degeneracy by assuming a field shape
of the form,
\begin{equation}
  F(M_D; M_{\mbox{\scriptsize{Dc}}},\sigma) =
\exp[-\case{1}{2}(M_D-M_{\mbox{\scriptsize{Dc}}})^2/\sigma^2], \label{gauss}
\end{equation}
where $M_{\mbox{\scriptsize{Dc}}}$ specifies the location and $\sigma$
specifies its width. This gaussian profile allows us to pinpoint the
location of the required magnetic field, by using observations of cyclic
variations of irradiance and effective temperature.

We either fix $\gamma$, or use
\begin{equation}
  \gamma(M_D) = 1+(B/B_m)^{1/5}(R_Z/200), \label{gauss2}
\end{equation}
to express the spatial and temporal variations of the direction of magnetic
fields. This implies that the more intense the magnetic field, the larger
its toroidal component. The profile of $\gamma$ is arbitrary, but its value must
lie between 1 and 2. Although the detailed results are sensitive to the 
functional form of $\gamma$, the qualitative features are same. We find that 
the magnetic effects are maximized when $\gamma=2$, so we set $\gamma=2$ in 
subsequent  computations. The main effect of a smaller $\gamma$, is to 
increase the magnetic field required to reproduce the observed variation 
of the sun.

As the magnetic fields that cause the solar activity
cycle are believed to exist outside of the solar core (where the
thermonuclear reactions occur), we do not consider the influence of
magnetic fields on the energy generation rate.

Magnetic fields affect the equation of state in the following way. If $\beta$,
the ratio of the gas pressure to the total pressure when magnetic fields are
present, is defined as
\begin{equation}
  \beta = 1 - \frac{\case{1}{3}aT^4}{P_T} - \frac{(\gamma-1)\chi\rho}{P_T},
\end{equation}
($a$ is the radiation constant), then the density of a gas can be determined
from
\begin{equation}
   \beta P_T = {\cal R}\rho T (1+E)/\mu_a.
\end{equation}
${\cal R}$ is the gas constant, $E$ is the number of free electrons
per nucleus (determined by solving the Saha equation) and $\mu_a$ is
the mean atomic weight per atom (Prather 1976). Note, $E$ also depends on
$\beta$. In the stardard case, since $\chi=0$, $\beta$ does not depend on 
$\rho$. But in the nonstandard case discussed in this paper, $\beta$ 
depends on $\rho$. Consequently, not only $\rho$, but also its first 
derivatives $\alpha$, $\delta$, $\nu$, and $\nu'$, and its second 
derivatives with respect to $T$ and $P_T$, all need to be calculated by 
iteration (which is  tedious). In the original implementation of this 
method by Lydon and Sofia (1995). The second derivatives were
approximated by assuming a perfect gas, which needs no iteration. In
this code upgrade, we modify the equation of state routine so that we can
accurately calculate all these thermodynamic variables by iteration. Although this
improves the numerical precision, it does not make any qualitative
difference. $\nabla_{\chi}$ and $\nabla_{\gamma}$ (defined in Appendix~\ref{appb}) are
also calculated by numerical derivatives, as are those time derivatives which
appear in the equations of stellar structure.

The qualitative difference comes from $f'$ (see Appendix \ref{appb}), 
which represents the influence of magnetic field on radiative loss 
of a convection element. The convection temperature gradient is now given by
\begin{equation}
  \nabla = \nabla_{\mbox{\scriptsize{ad}}} + (y'/V'\gamma_0^2C)(1+y'/V')-A_m,
\end{equation}
where
\begin{equation}
   A_m = f' [(\nu/\alpha)\nabla_\chi +
(\nu'/\alpha)\nabla_\lambda]\nabla_{\mbox{\scriptsize{ad}}}. \label{am1}
\end{equation}
All symbols used here are defined in Appendix~\ref{appb}. When $f'=1$ and $\gamma$ is fixed, 
equation~(\ref{am1}) reduces to 
\begin{equation}
A_m = (\nu/\alpha)\nabla_\chi\nabla_{\mbox{\scriptsize{ad}}}.
\label{am0}
\end{equation}
However, under those assumptions, the calculated cyclic variation
of effective temperature is in antiphase with the solar activity, contrary to the
observations \cite{GL97b}. In order to get a proper fit to the observations, it is 
necessary to assume $f'=3$, a value that will be used in all our subsequent 
calculations.

We use Eq.~(\ref{kappap}) to compute the global magnetic effect on the radiative 
opacity, and find that this has little influence on the effective temperature.
On the other hand, the need of $f'=3$ implies that the energy loss of a 
convective element due to turbulence associated with 
the local magnetic field, is much more efficient than the radiative loss 
enhancement of the convective element due to the global magnetic field.
 
\vspace{3mm}
\centerline{\epsfysize=5.5cm \epsfbox{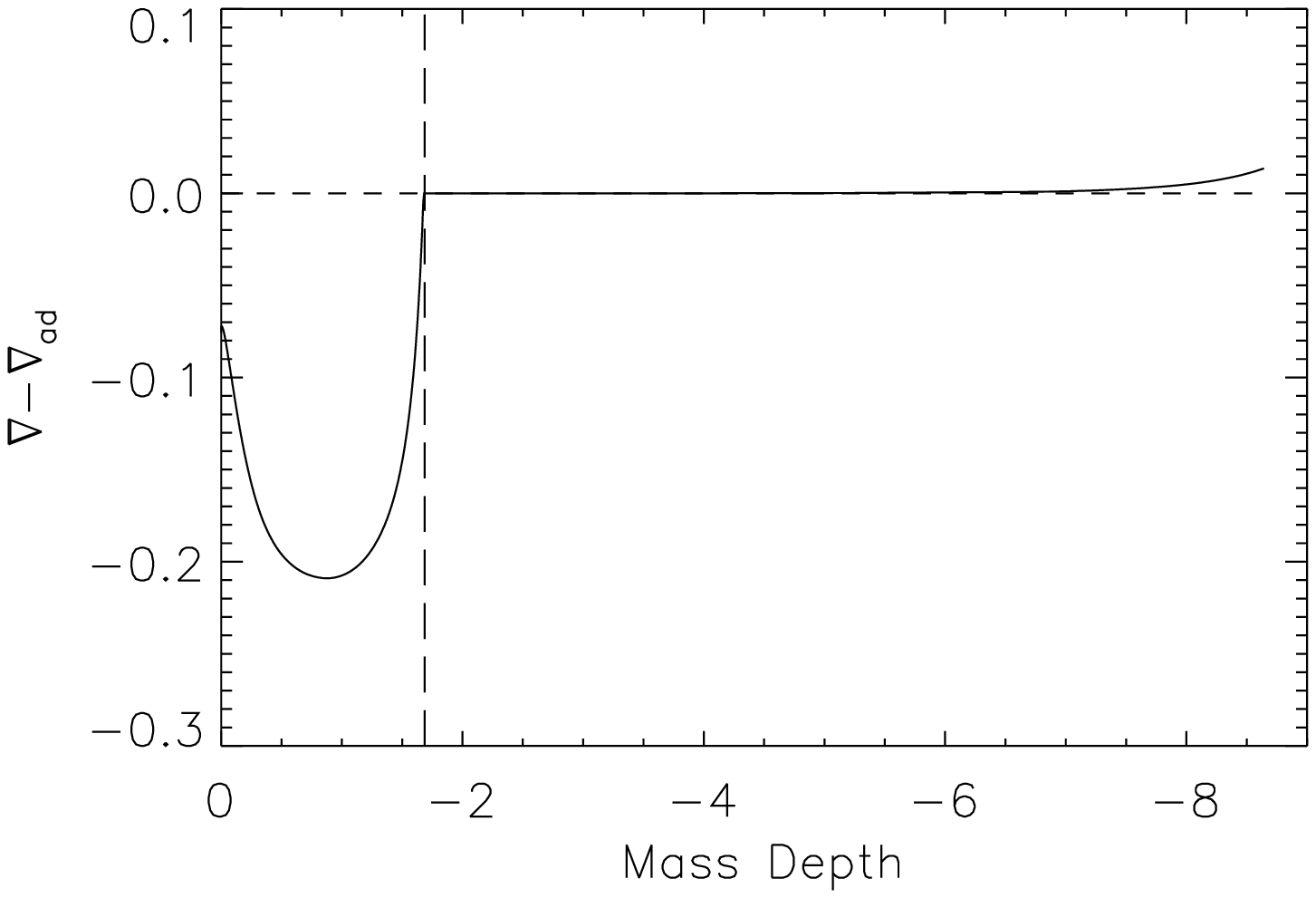}}
\figcaption[lifig3.eps]{
The difference between the temperature gradient and the adiabatic gradient
as a function of the mass depth in the solar interior. The dashed line marks
the location of lower boundary surface of the convection zone.
\label{lifig3}
}
\vspace{3mm}

We do not redistribute the mass grid points, nor do we predict stellar
structure variables for the next time step, when the timestep is smaller
than a solar cycle period, since these operations will generate an error
much larger than $10^{-10}$ per year for luminosity. All convergence
criteria for iteration are tightened by requiring the relative error to be
no larger than $6\times 10^{-16}$ or so. In this modified code, the matching
point is at $M_D\approx -8.6$.

The conventional model parameters are, the mixing length parameter (the ratio
of the mixing length to the pressure scale height, usually denoted as
$\alpha$ parameter), the initial hellium mass abundance ($Y_0$), and the initial
heavy element mass abundance ($Z$). The magnetic parameters introduced in
this paper are the peak mass depth $M_{\mbox{\scriptsize{Dc}}}$ of the
magnetic energy density $\chi$, the width of $\chi$ (i.e.,$\sigma$), the
magnetic strength parameter $B_0$, and the turbulent loss parameter of energy
of a convective element $f'$. It is well known that the conventional
parameters can be determined by present observations of solar radius and 
luminosity. Through numerical experiments (i.e., model calculations), we find
that the magnetic parameters ($M_{\mbox{\scriptsize{Dc}}}$, $\sigma$, $B_0$
and $f'$), can be determined by the observed cyclic variation of solar
radius, luminosity and effective temperature. We fix $f'=3$ in all 
subsequent runs, as required by the observed variation of effective
temperature. $\gamma=2$ is fixed to maximize the magnetic effects.
In order to probe the observation-required location, strength and profile of
the solar interior magnetic field, we change 
$M_{\mbox{\scriptsize{Dc}}}$, $\sigma$ and $B_0$ from run to run.

\section{Constraints on the solar interior magnetic field}

\subsection{The overshoot layer}

At present, it is generally believed that the magnetic field is stored only
in the subadiabatically stratified overshoot layer in the form of flux
tubes, which may break loose due to kink instabilities, when a threshold of
$10^5$ gauss is exceeded \cite{CMS95,CMS98}. Figure~\ref{lifig3} depicts
$\nabla-\nabla_{\mbox{\scriptsize{ad}}}$ vs $M_D$, in which the
subadiabatically stratified overshoot layer approximately corresponds to the
mass depth range: $-1.68\le M_D<-1.2$. $M_{\mbox{\scriptsize{Dc}}} = -1.68$
marks the base of the convection zone of our model sun. Therefore,
the maximum yearly-averaged field strength $B_m$ in the overshoot layer
during the solar cycle should be smaller than the threshold. The field with
$M_{\mbox{\scriptsize{Dc}}}=-1.45$ and $\sigma<0.2$, can be considered to be
confined in the overshoot layer. However, even if the threshold value of B
is used, the resultant irradiance and effective temperature variations are
negligible.

Only when we abandon this threshold constraint (e.g., using a field larger
than $2.6\times10^6$ gauss), can we get irradiance variations comparable to
the observations, in magnitude and phase; however, even then, the effective
temperature variation is much smaller and in antiphase with the solar cycle.
Besides, the resulting relative radius variation is larger than $0.03\%$.
Since the observations reveal that both irradiance and effective temperature
variations are in phase with the solar cycle, and the relative radius
variation should be much smaller than $0.01\%$ \cite{EKBS00}, we conclude
that the magnetic field that gives rise to the observed variations of 
irradiance and effective temperature (shown in Figures~\ref{lifig1} 
and \ref{lifig2}) cannot be completely confined in the overshoot layer.

\subsection{The convection zone}

The above arguments lead us to investigate the possibility that the magnetic 
field is located totally within the convection zone. We set $\sigma=0.5$ in 
this region. In order to reproduce the observed cyclic variation of irradiance
and effective temperature, $B_0$ will increase if $\sigma$ decreases, and vice 
versa. We choose $M_{\mbox{\scriptsize{Dc}}}$ from the interval (-8.6,-1.68], 
which covers the convection zone of our numerical solar model. For each 
selected value for $M_{\mbox{\scriptsize{Dc}}}$, we change $B_0$ in order 
to reproduce the observed cyclic variations of solar irradiance and effective 
temperature by using our solar evolution code.

\subsubsection{The lower part}

We find that the lower part of the convection zone, 
$-1.68\ge M_{\mbox{\scriptsize{Dc}}}>-4.2$, is ruled out because 
it produces radius changes that are too large ($\Delta\ln R >0.01\%$) 
and/or because it requires a too strong magnetic field 
($B>5\times 10^4$ G).

\subsubsection{The intermediate part}

For the next region up, $-4.2>M_{\mbox{\scriptsize{Dc}}}>-7.8$, or $0.959
R_{\sun} < R  < 0.997 R_{\sun}$, we find that our model can reproduce the 
observed cyclic variations of irradiance and effective temperature by using
a possible equipartition magnetic field, $B_m < 5\times\times10^4$ G.
In this case, the amplitude of the radius changes is from 17 to 3 km, 
and $W=\Delta\ln R/\Delta\ln L$ ranges from 0.02 to 0.006.

In order to pinpoint the location of the required magnetic fields in the
allowed region of solar convection zone
($M_{\mbox{\scriptsize{Dc}}}\in[-7.8,-4.2]$), we need more information. We
obtain the allowed region by using the observational information for solar
irradiance and effective temperature cyclic variations, and by assuming the
corresponding solar radius variation $\Delta\ln R$ to be much smaller than
$0.01\%$. If we have the direct observational information for $\Delta\ln R$
or the $W$ parameter, we can pinpoint the location of solar interior magnetic
fields. Since we do not have simultaneous observational information for the
radius cyclic variation in the period ranging from 1978 to 1992, we must
invoke some physical considerations.

\vspace{3mm}
\centerline{\epsfysize=5.5cm \epsfbox{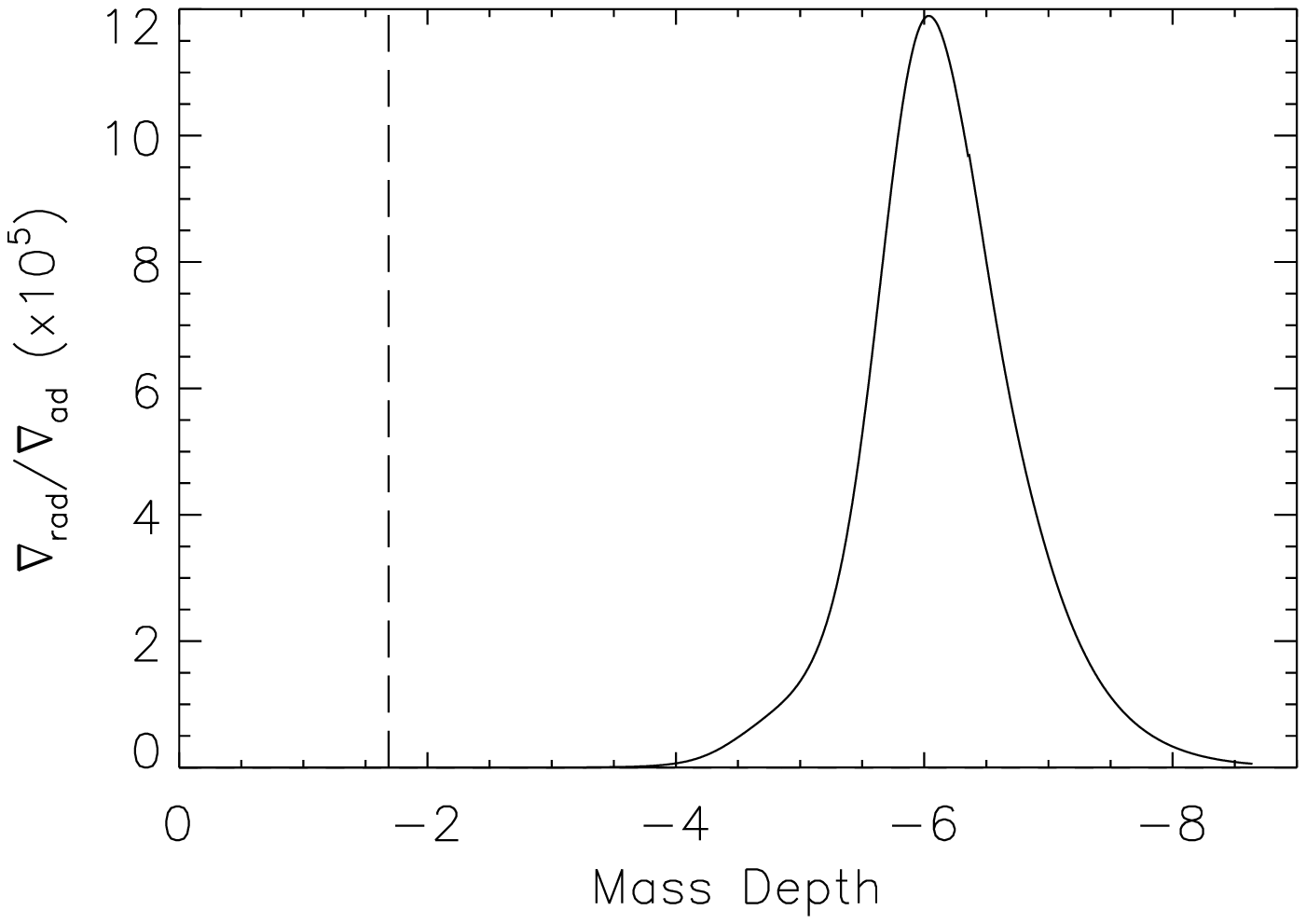}}
\figcaption[lifig4.eps]{
The ratio of radiative to adiabatic temperature gradients as a function of
mass depth in the solar interior. The dashed line marks the location of
lower boundary surface of the convection zone.
\label{lifig4}
}
\vspace{3mm}

Figure~\ref{lifig4} shows the ratio of radiative to adiabatic temperature
gradients as a function of mass depth in the sun, from which we can see that
the layer with $M_D$ ranging from $-4$ to $-8$ is the most unstable
convective region. In this layer superstrong plasma turbulence is inevitable,
since the radiative gradient $\nabla_{\mbox{\scriptsize{rad}}}$ is much
larger than the adiabatic temperature gradient
$\nabla_{\mbox{\scriptsize{ad}}}$. Therefore, this may suggest a turbulence
generation mechanism for solar magnetic fields. Strong plasma turbulence can
generate strong small-scale magnetic fields, which form small magnetic cells
with almost random orientation. The residual field of the previous cycle and
differential rotation, tend to align and bundle these random magnetic
``needles'' to form flux tubes.

The solid curve in Figure~\ref{lifig5}  (choosing
$M_{\mbox{\scriptsize{Dc}}}=-6.5$, $\sigma=0.5$, $B_0=7$ G) shows the
corresponding magnetic field distribution in the solar interior in 1989,
near solar maximum. Figure~\ref{lifig6} compares the calculated (solid
curve) and measured (dash-dotted curve) cyclic irradiance variation, while
Figure~\ref{lifig7} compares the calculated (solid curve) and measured
(dash-dotted curve) cyclic effective temperature variation. The irradiance
fit ($\chi^2 = 2$) is better than the effective temperature fit ($\chi^2=
47$). The computed radius variation is $5\times10^{-6}$, and the $W$
parameter is almost constant (about $4\times10^{-3}$). Figure~\ref{lifig8}
shows the calculated structure change generated by the field in 1989 
(solid curve).

\vspace{3mm}
\centerline{\epsfysize=5.5cm \epsfbox{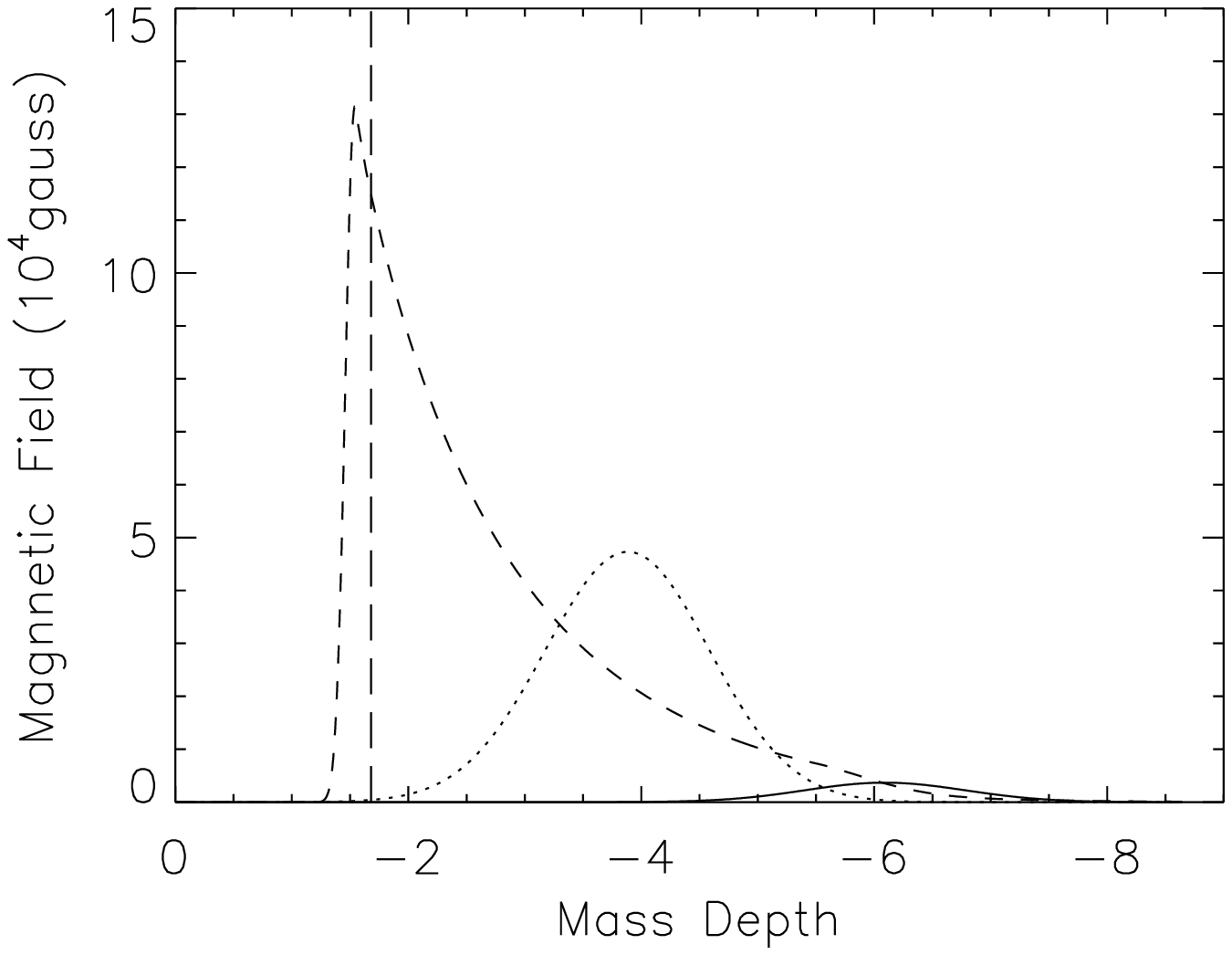}}
\figcaption[lifig5.eps]{
Three possible distributions of inferred magnetic field in the solar
interior in 1989 according to the measured irradiance and photospheric
temperature cyclic variations given in Figures~\ref{lifig1}
and~\ref{lifig2}. The vertical line indicates the base of the convection
zone.
\label{lifig5}
}
\vspace{3mm}

Although we can relate the magnetic field generation to the plasma
turbulence in this region, the magnetic field maintenance here faces the old
problem of magnetic buoyancy. How can we get around this problem? 
Observations and simulations suggest that the actual dynamics within 
solar convection zone is extremely intricate
\cite{BCT95,KC98,N99}. The velocities and magnetic fields are complex,
exhibiting large-scale structure and ordered behavior amidst rapidly varying
and intense small-scale turbulence. In fact, the numerical simulations of
the solar outer convection zone \cite{KC98,N99} indicate a major presence of
downward moving plumes. These ordered downdrafts may gather small magnetic
flux tubes, generated in the extremely unstable layer, to form larger flux
tubes and carry them to the deeper layer. These downdrafts may also push
down the magnetic flux tubes to balance the magnetic buoyancy to form a
magnetized layer below the most unstable convective region.

One possibility is that the downdrafts gather and carry magnetic flux tubes,
generated by turbulence, to depths in the convective envelope until some
sort of equipartition is reached. In fact, Antia, Chitre, and Thompson
(2000) have explored this possibility. They employ the observed splittings
of solar oscillation frequencies to separate the effects of interior solar
rotation, and to estimate the contribution from a large-scale magnetic field.
After subtracting out the estimated contribution from rotation, there is
some residual signal in the even splitting coefficients. This may be
explained by a magnetic field of approximately 20 kG strength located at a
depth of $M_D=-4.25$ ($r=0.96 R_{\sun}$) in 1996. Since the density near
$M_D=-4.25$, is of order $4\times10^{-3}$, and the downward velocity for
the plumes is of order $5\times10^4$ cm s$^{-1}$, the estimated dynamical
pressure of the plumes, $\rho v^2$, is equal to or larger then $10^7$ dyne
cm$^{-2}$. The size of $\rho v^2$ is comparable with the magnetic 
pressure, $B^2/8\pi$, corresponding to a field strength of $20-30$ kG.
This demonstrates that a
stable magnetized layer in the convection zone proper, may form when the
complexity of convection motion is taken into account. If we wish to
reproduce the observed temperature and irradiance variations by the magnetic
field at the depth indicated by helioseismology, we find that $B_m$ ranges
from 20 kG to 47 kG during a solar cycle (choosing $B_0=90$ G). Therefore,
in Figures~\ref{lifig5}-\ref{lifig8} we also show this case (dotted curves:
$M_{\mbox{\scriptsize{Dc}}}=-4.25$, $\sigma=0.5$). The irradiance fit
($\chi^2 = 1$) is also better than the effective temperature fit ($\chi^2=
43$), the $W$ parameter is equal to about $2\times10^{-2}$, and the
predicted cyclic variation of solar radius is equal to about
$2\times10^{-5}$, which can be tested by measuring the $W$ parameter.

In fact, using MDI/SOHO data obtained between April 19, 1996 and June 24,
1998, Emilio et al (2000) found $W\le2\times10^{-2}$, which is consistent
with the above prediction. This shows that the second case is in agreement 
with all relevant precise observations, including solar irradiance, effective
temperature, radius, and p-mode oscillation observations. The first case is
ruled out by the $W$ parameter inferred from the observation.

\subsubsection{The upper part}

The upper part of the convection zone, $M_{\mbox{\scriptsize{Dc}}}<-7.8$, 
is ruled out because the resulting irradiance and effective temperature 
variations are in antiphase with the solar cycle, which is in conflict with
the observations.

In all cases, to produce the observed luminosity and effective temperature
variations, the required magnetic field must increase with depth. The
resulting $W$ parameter also increases with depth. This is in agreement with
the early studies \cite{D83,EST85}.

\subsection{The extended layer}

Another possibility is that these magnetic flux tubes are carried downwards
into the stable subadiabatically stratified overshoot layer, to provide a
seed field for the dynamo operating at the base of the convection zone. Therefore,
the magnetic field extends from the convection zone to the overshoot layer. The
following magnetic field distribution mimics this case,
\begin{equation}
  F= \left\{ \begin{array}{lll} h(700)   & \mbox{ $M_D\ge
M_{\mbox{\scriptsize{Dc}}}$} \\
                  h(1)  - 0.325(M_D-M_{\mbox{\scriptsize{Dc}}}) &  \mbox{
$M_D\le M_{\mbox{\scriptsize{Dc}}}$} \\
                    h(1) + 1.3\exp[-2(M_D+5.55)^2] & \mbox{ $M_D\le -5.55$}
\end{array} \right.
    \label{step}
\end{equation}
where $h(m)= (M_D/M_{\mbox{\scriptsize{Dc}}})^m
\exp[-m(M_D/M_{\mbox{\scriptsize{Dc}}}-1)]$. Choosing
$M_{\mbox{\scriptsize{Dc}}}=-1.55$ in equation~(\ref{step}) and $B_0=250$ G
in equation~(\ref{bm}), we also get a good fit to
Figures~\ref{lifig1} ($\chi^2=1$) and \ref{lifig2} ($\chi^2=40$). The
magnetic field profile is depicted as a dashed curve in Figure~\ref{lifig5}.
This case is also represented by dashed curves in
Figures~\ref{lifig6}-\ref{lifig8}. The predicted $W$ parameter is about
$2\times10^{-2}$. It should be noted that what contributes to the observed 
cyclic variations of solar irradiance and effective temperature
in this case, is the field located within the convection zone proper, while the
contribution of the field confined in the overshoot layer is negligible.
MDI/SOHO data do not rule out this possibility.

\vspace{3mm}
\centerline{\epsfysize=5.5cm \epsfbox{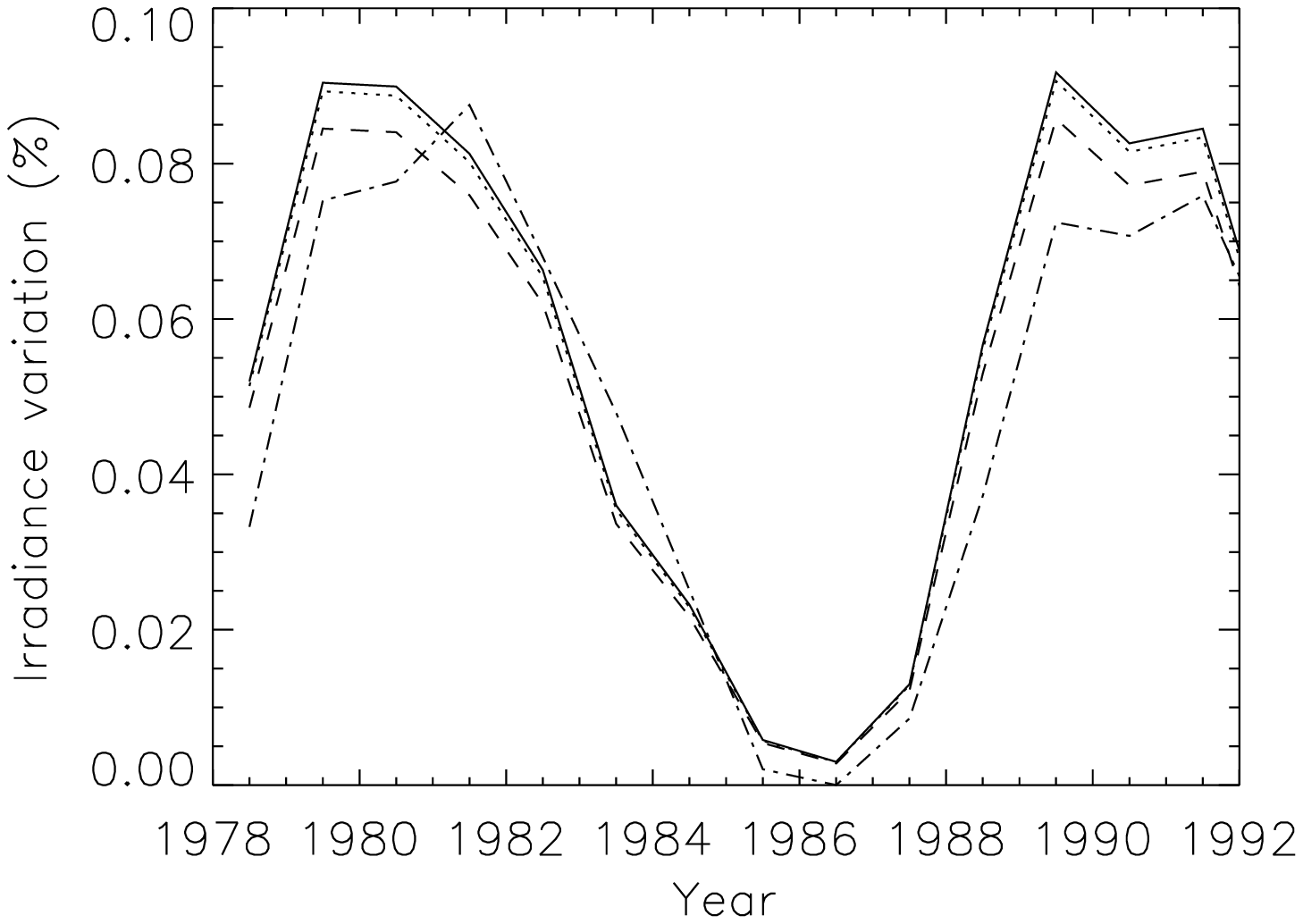}}
\figcaption[lifig6.eps]{
Comparison between the measured (dot-dashed curve) and calculated solar 
irradiance variations. The line style for the calculated is the same as that
in Fig.~\ref{lifig5}.
\label{lifig6}
}
\vspace{3mm}

\vspace{3mm}
\centerline{\epsfysize=5.5cm \epsfbox{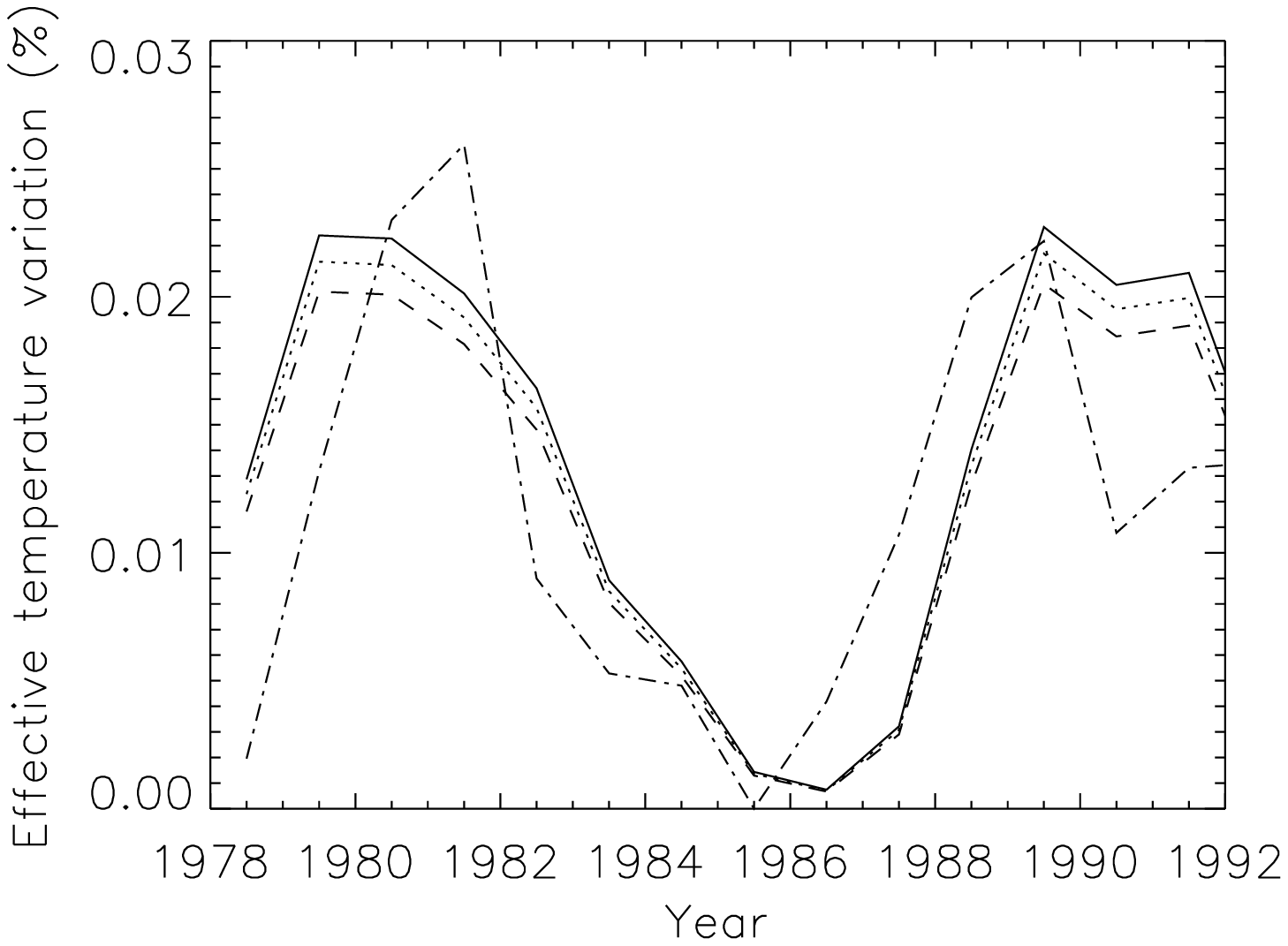}}
\figcaption[lifig7.eps]{
Comparison between the measured (dot-dashed curve) and calculated 
solar photospheric temperature variations. The line style for the 
calculated is the same as that in Fig.~\ref{lifig5}.
\label{lifig7}
}
\vspace{3mm}

\section{Conclusions}

It is possible to locate solar interior magnetic fields using the observed
cyclic variations of three global solar parameters such as luminosity,
temperature and radius at the surface of the sun. This provides an
alternative to helioseismology as a probe of the solar interior magnetic
fields. Simultaneous measurements of solar total irradiance and effective
temperature, can only select an allowed range of solar internal magnetic
fields, which in terms of magnitude and location are consistent with 
helioseismic observations and recent MDI experiment on SOHO.

Although the observed cyclic variations of solar irradiance, effective
temperature, radius and p-mode oscillation frequencies, require a magnetic
field component between 20 and 47 kG, peaked at $r=0.96R_{\sun}$ (within the
convection zone proper), a stronger component of about 300 kG buried in the
overshoot layer beneath the base of the convection zone cannot be ruled out,
since the contribution of the latter to those observations is negligible.

\vspace{3mm}
\centerline{\epsfysize=9.5cm \epsfbox{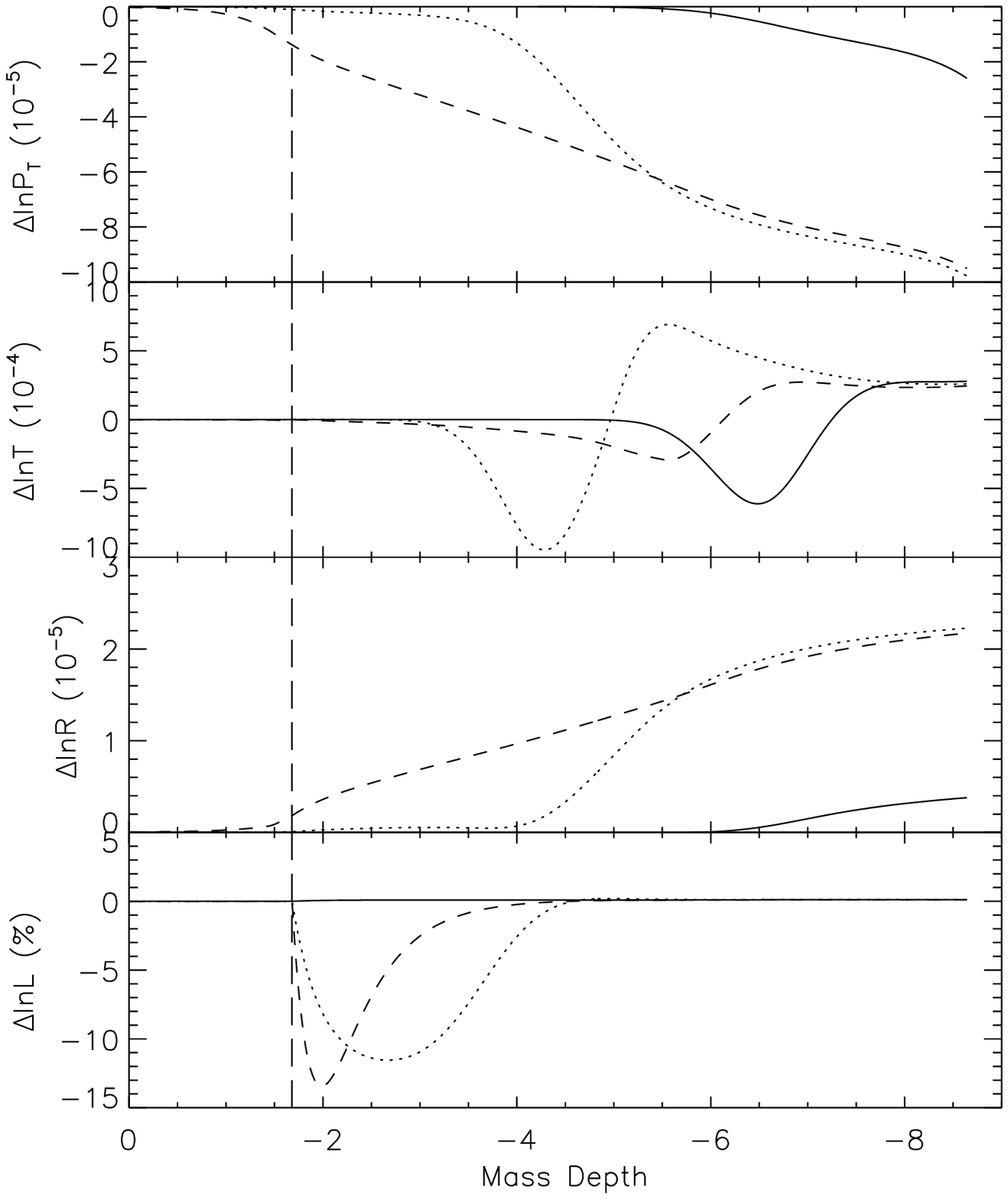}}
\figcaption[lifig8.eps]{
The structural changes caused by the magnetic field distributions given in
Figure~\protect\ref{lifig5}: relative pressure, temperature, radius and
luminosity changes from top to bottom. The vertical line indicates the base
of the convection zone. The line style for the calculated is the same as that
in Fig.~\ref{lifig5}.
\label{lifig8}
}
\vspace{3mm}

\acknowledgments

This work was supported in part by a grant from the National Aeronautics and Space Administration, and in part by Natural Science Foundation of China (project 19675064).

\appendix

\section{An estimate of magnetic effects on the radiative
opacity}\label{appa}

A magnetic field affects the absorption processes associated with
free electrons, such as electron scattering absorption
$\kappa_{\mbox{\scriptsize{sc}}}$ and free-free transition absorption
$\kappa_{\mbox{\scriptsize{ff}}}$. We can estimate this effect in terms of
the mean free paths with and without a magnetic field, $l_B$ and $l_0$
\begin{equation}
   \kappa_{\mbox{\scriptsize{e0}}}/\kappa_{\mbox{\scriptsize{eB}}} \propto
(l_B/l_0)^2 \approx 1
     +\case{1}{4} (\gamma-1)\tau^2_e \Omega^2_e \label{kappa}
\end{equation}
Here $1/\tau_e$ is the collision frequency between electrons,
$\kappa_e=\kappa_{\mbox{\scriptsize{sc}}}+\kappa_{\mbox{\scriptsize{ff}}}$
is the absorption component that will be affected by magnetic field and
$\Omega_e$ is the electron cyclotron frequency.
Magnetic fields also affect the bound-bound absorption, but this effect
cannot be estimated easily by means of the classic method. Nevertheless, the
quantum effect must be much smaller than the classic one described above.

The influence of magnetic fields on the radiative opacity can only be
treated approximately, since we only use opacity tables. We assume that we
can decompose the total opacity coefficient $\kappa$ into two parts:
$\kappa=\kappa_e + \kappa_1$, since $\kappa_{\mbox{\scriptsize{sc}}}$ and
$\kappa_{\mbox{\scriptsize{ff}}}$ do not or only weakly depend on frequency
(before taking the Rosseland mean). We use the opacity subroutine provided
by YREC7 to calculate $\kappa$, and use equations~(17.2) and (17.5) in
Kippenhahn and Weigert (1990) to calculate $\kappa_e$. The opacity corrected
by magnetic fields, $\kappa'$, can be expressed approximately as follows
\begin{equation}
  \kappa'=\kappa -
\kappa_e\frac{(\gamma-1)\tau_e^2\Omega_e^2}{4+(\gamma-1)\tau_e^2\Omega_e^2}.
\label{kappap}
\end{equation}
The approximation originates from the fact that the absorption due to
free-free transitions depends on frequency as  $\kappa_\nu \propto
\nu^{-3}$.

\section{Convection temperature gradient}\label{appb}

When both $\chi$ and $\gamma$ are considered to be variables, the convection
temperature gradient $\nabla = (\partial \ln T/\partial P_T)_s$ (s stands
for surroundings) can be expressed as follows
\begin{equation}
  \nabla = \nabla_{\mbox{\scriptsize{ad}}} + (y'/V'\gamma_0^2C)(1+y'/V') -
A_m,
\end{equation}
where the initial $\chi$ of the convection element is assumed to remain
frozen in the surrounding  material.
$\nabla_{\mbox{\scriptsize{ad}}}=(\partial \ln T/\partial \ln P_T)_S$ ($S$
stands for entropy) is the adiabatic gradient and $y'$ is obtained by solving
the cubic algebraic equation,
\begin{equation}
   0 = 2A{y'}^3 +V'{y'}^2 +{V'}^2y'-V'.
\end{equation}
$\gamma_0$, $C$, $V'$, $A_m$ and $A$ are defined by
\begin{eqnarray}
   \gamma_0 &=& [(c_{\rm p}
\rho)/(2acT^3)][(1+(1/3)\omega^2/\omega], \\
   C &=& (g/l_m^2\delta)/8H_p, \\
   A_m &=& f' [(\nu/\alpha)\nabla_\chi +
(\nu'/\alpha)\nabla_\lambda]\nabla_{\mbox{\scriptsize{ad}}}, \label{am} \\
   V'&=& 1/[\gamma_0
C^{1/2}(\nabla_{\mbox{\scriptsize{rad}}}-\nabla_{\mbox{\scriptsize{ad}}}+A_m)^{1/2}], \\
   A &=& (9/8)[\omega^2/(3+\omega_2)].
\end{eqnarray}
$\omega=\kappa \rho l_m$, $g$ is the gravity acceleration, $H_p$ is the
pressure scale height, $l_m$ is the mixing length and
$\nabla_{\mbox{\scriptsize{rad}}}$ is the radiative temperature gradient.
$f'$ is a dimensionless parameter that determines the influence of magnetic
field on radiative loss of a convective element, $\nabla_{\chi} = (\partial
\ln\chi/\partial\ln P_T)_s$, and $\nabla_{\gamma} = (\partial\ln
\gamma/\partial \ln P_T)_s$. In general, magnetic fields tend to inhibit
convection.

\end{document}